\pdfoutput=1
\documentclass[11pt]{article}

\usepackage{EMNLP2023}

\usepackage{times}
\usepackage{latexsym}

\usepackage[T1]{fontenc}
\usepackage[utf8]{inputenc}

\usepackage{microtype}

\usepackage{inconsolata}

\usepackage{listings}
\usepackage{booktabs}
\usepackage{graphicx} % Needed for including graphics
\usepackage{color}
\definecolor{codegreen}{rgb}{0,0.6,0}
\definecolor{codegray}{rgb}{0.5,0.5,0.5}
\definecolor{codepurple}{rgb}{0.58,0,0.82}
\definecolor{backcolour}{rgb}{0.80, 0.92, 0.95}
\lstdefinestyle{mystyle}{
    backgroundcolor=\color{backcolour},   
    commentstyle=\color{codegreen},
    keywordstyle=\color{magenta},
    numberstyle=\tiny\color{codegray},
    stringstyle=\color{codepurple},
    basicstyle=\footnotesize,
    breakatwhitespace=false,         
    breaklines=true,                 
    captionpos=b,                    
    keepspaces=true,                 
    numbers=left,                    
    numbersep=5pt,                  
    showspaces=false,                
    showstringspaces=false,
    showtabs=false,                  
    tabsize=2
}
\title{PyTerrier-GenRank: The PyTerrier Plugin for Reranking with Large Language Models}

\author{Kaustubh D. Dhole \\
  Department of Computer Science \\
  Emory University\\
  Atlanta, GA -- USA \\
  \texttt{\textcolor{darkblue}{kdhole@emory.edu}}
 }

\begin{document}
\maketitle
\begin{abstract}
\vspace{-5pt}
Using LLMs as rerankers requires experimenting with various hyperparameters, such as prompt formats, model choice, and reformulation strategies. We introduce~\href{https://github.com/emory-irlab/pyterrier_genrank}{\underline{\textbf{PyTerrier-GenRank}}}, a PyTerrier plugin to facilitate seamless reranking experiments with LLMs, supporting popular ranking strategies like pointwise and listwise prompting. We validate our plugin through HuggingFace and OpenAI hosted endpoints.
\end{abstract}

\section{Introduction}

Generative reranking methods have recently gained popularity for their ability to produce a more optimal ranking of documents by prompting large language models. These methods are also employed in Retrieval Augmented Generation~\cite{dhole-2024-kaucus} paradigms as they are effective for enhancing generation through the incorporation of retrieved information. Typically, a sparse retriever such as BM25~\cite{robertson1992okapi} or a dense retriever like ColBERT~\cite{khattab2020colbert} is utilized to retrieve a small subset (ranging from 20 to 1000) from among millions or billions of text pieces. This subset is then forwarded to a second-stage generative reranker for reranking.

The increasing adoption of such reranking methods necessitates rigorous testing of various hyperparameters, including employing different prompts, evaluating performance with different model weights, and using it in conjunction with various retrieval, reranking, and reformulation~\cite{dhole2024queryexplorer, dhole2024generative} pipelines.

In that regard, we introduce --~\textbf{PyTerrier-GenRank} -- to encourage rapid experimentation of retrieval pipelines employing recently popular LLM-based reranking methods. The code includes PyTerrier~\cite{pyterrier2020ictir} wrappers over many helper functions of the RankLLM repository~\cite{pradeep2023rankvicuna, pradeep2023rankzephyr}. Our plugin can be quickly installed in any Python environment and is available~\href{https://github.com/emory-irlab/pyterrier_genrank}{\textcolor{codegreen}{at this link}}\footnote{\href{https://github.com/emory-irlab/pyterrier_genrank}{\textcolor{codegreen}{github.com/emory-irlab/pyterrier\_genrank}}}.

\label{Listing1}
\begin{lstlisting}[language=Python, caption=Retrieving using a listwise reranker]
 
 import pyterrier as pt
 from rerank import LLMReRanker

 bm25 = pt.BatchRetrieve(dataset.get_index(), wmodel="BM25")

 reranker = LLMReRanker("castorini/rank_vicuna_7b_v1")

 pipeline = bm25 % 100 
    >> pt.text.get_text(index, 'text') 
    >> reranker

 pipeline.search('Indian restaurants')
 
\end{lstlisting}
\section{Generative Reranking}
Large Language Models (LLMs) have been widely used for reranking tasks in various ways involving novel hyperparameters. Among the most popular ones have been ranking documents in a pointwise, pairwise and listwise fashion. MonoBERT~\cite{monobert} and MonoT5~\cite{nogueira-etal-2020-document} are among the earliest pointwise ranking methods that were trained to emit a single scalar relevance score for a query-document pair. In contrast, pairwise approaches~\cite{qin-etal-2024-large} involved prompting models with a query and two documents to compare and rank them.~\texttt{RankVicuna}~\cite{pradeep2023rankvicuna},~\texttt{RankZephyr}~\cite{pradeep2023rankzephyr}, and~\texttt{RankGPT}~\cite{sun2023chatgpt, openai2023gpt35} have explored listwise ranking strategies by prompting their respective models with a list of documents and generating a ranked list of document IDs. These paradigms have also been investigated in the context of recommender systems~\cite{yue2023llamarec}.

The number of hyperparameters involved in creating an optimal retriever, the choice of prompts, the number of documents, the weight of the reformulation strategy, etc., all influence the final ranking set in various ways. Besides, the increasing number of LLMs released by various organizations further necessitates rigorous testing to evaluate performance across different settings. To ensure the validity and reproducibility of results, it is imperative that researchers and developers can easily replicate experiments across these novel sets of hyperparameters. 

\section{Python IR Toolkits}
We choose PyTerrier~\cite{pyterrier2020ictir} due to its modularity and Python-friendly interface. It is built on the Terrier search engine and excels in modular, rapid experimentation for IR research, allowing users to easily integrate and evaluate various retrieval and re-ranking models. It has been employed in various studies to evaluate retrieval effectiveness~\cite{dhole2024genqrensemble, datta2024deep, parry2024top, qppdutta, dhole2024llmjudge}.
\begin{table}[h!]
\centering
\resizebox{\columnwidth}{!}{
%\begin{tabular}{>{\raggedright\arraybackslash}p{1.2cm} >{\raggedright\arraybackslash}p{2cm} >{\raggedright\arraybackslash}p{2cm}}
\begin{tabular}{llc}
\toprule
\hline
\textbf{Training} & \textbf{Name} & \textbf{nDCG@10} \\
\hline
\midrule
0-shot & BM25 & .480 \\
0-shot & gpt-35-turbo$^{\beta}$~\cite{openai2023gpt35} & .660 \\
0-shot & gpt-4-turbo-0409$^{\beta}$~\cite{gpt4turbo0409} & .701 \\
0-shot & gpt-4o-mini~\cite{gpt4omini} & \textbf{.710} \\
0-shot & Llama-3-8B-Instruct~\cite{llama3modelcard} & .572 \\
0-shot & Llama-3.1-8B-Instruct~\cite{llama3_1modelcard} & .594 \\
0-shot & Llama-Spark (8B)~\cite{spark} & \textbf{.612} \\
\hline
Trained & Rank-Vicuna$^{\beta}$~\cite{pradeep2023rankvicuna} & .672 \\
Trained & Rank-Zephyr$^{\beta}$~\cite{pradeep2023rankzephyr} & .711 \\
\hline
\bottomrule
\end{tabular}
}
\caption{Comparing reranking performance of different models for TREC-DL 2019~\cite{craswell2020overview} using PyTerrier\_Genrank}
\label{tableres}
\end{table}

\section{Generative Reranking Wrappers}
Our plugin employs PyTerrier wrappers over reranking class functions modified from the Pyserini~\cite{lin2021pyserini} based~\href{https://github.com/sunnweiwei/RankGPT}{RankGPT} and~\href{https://github.com/castorini/rank_llm}{RankLLM} repositories. We allow additional hyperparameters to be expressed from function calls. We provide sample code to execute HuggingFace~\cite{wolf2020transformers} models and OpenAI-based endpoints.

\section{Sanity Check}
As a sanity check, we re-evaluate some previously published rerankers like RankVicuna, RankGPT, using this framework. Those marked with superscript $^{\beta}$ have already been previously tested with the Pyserini framework in their respective papers. We also evaluate the zero-shot retrieval effectiveness of newly published instruction-tuned models like LLama-Spark.

We find that among the 8B zero-shot approaches, LLama-Spark is the most effective as a reranker. The complete results are shown in Table~\ref{tableres}.

\section*{Ethics Statement}
Large Language Models are embedded within a broader socio-technical framework~\cite{dhole2023large,dhole2023nl}. When deploying them for tasks such as ranking and recommendations, it's crucial to assess their long-term impacts, including the potential for creating echo chambers and the risk of generating biased suggestions.

\bibliography{anthology,custom}
\bibliographystyle{acl_natbib}

\appendix

\end{document}